\documentclass[fleqn,twoside]{article}
\usepackage{espcrc2}

\usepackage{graphicx}
\usepackage[figuresright]{rotating}

\newcommand{\AmS}{{\protect\the\textfont2
  A\kern-.1667em\lower.5ex\hbox{M}\kern-.125emS}}

\title{High Energy Neutrino Cross Sections}

\author{Mary Hall Reno\address{
        Department of Physics and Astronomy,
        University of Iowa, Iowa City, Iowa 52242 USA}\thanks{This
        work is supported in part by D.O.E. contract DE-FG02-91ER40664.} }
\begin{document}

\begin{abstract}
The theoretical status of the neutrino-nucleon cross section is
reviewed for incident neutrino energies up to $E_\nu=10^{12}$ GeV,
including different approaches to high energy extrapolations.
Nonstandard model or nonperturbative physics may play a role at
ultrahigh energies. The cases of mini-black hole production and
electroweak instanton contributions are discussed as examples
in the context of ultrahigh energy neutrino scattering.

\vspace{1pc}
\end{abstract}

\maketitle

\section{INTRODUCTION}

Neutrinos from cosmic ray interactions in the atmosphere and
from solar fusion processes have already opened a new era of
neutrino physics as we study the masses and mixing parameters
of neutrinos \cite{solaratm}. Much higher energy neutrinos hold the promise
to reveal aspects of particle physics, but also to help us understand
mechanisms for cosmic acceleration and the environment in
astrophysical sources where high energy cosmic rays are produced \cite{lm}.

Neutrinos are unique in their properties of being undeflected by
magnetic fields, and for traversing astronomical distances unabsorbed.
Large under-water and under-ice experiments using Cherenkov
radiation are dedicated to detecting
neutrinos from astrophysical sources, and other experiments such
as air shower arrays and air fluorescence detectors have the
possibility
to detect neutrinos via horizontal air showers. Combined, these
experiments probe sources of neutrinos over a phenomenal range
of energies, potentially up to the highest energies seen
in cosmic ray experiments.

An essential ingredient for the interpretation of neutrino results
is the neutrino-nucleon cross section. We review here the status of the
neutrino-nucleon cross section up to $E_\nu=10^{12}$ GeV. Experimental
results from $ep$ scattering at HERA \cite{h1,zeus} make the standard
model neutrino-nucleon
cross section very well understood up to approximately $10^7$ GeV.
At higher energies, extrapolations of the parton distribution
functions
(PDFs) make for some uncertainties in the theoretical predictions.
We review the different approaches to the high energy extrapolations.

At the highest energies, one considers the possibility that non-standard
model physics may play a role in neutrino interactions. The
contribution
of mini-black holes to the neutrino-nucleon cross section is briefly
discussed as an example of how non-standard model physics comes into
play at high energies. In addition, there have been proposals that
standard model electroweak instantons may contribute significantly
to the ultrahigh energy neutrino cross section. These additions to
the standard model (perturbative) cross section are discussed
in the fourth section.

\section{STANDARD MODEL FORMALISM}

The differential cross section for neutrino scattering
with an isoscalar nucleon $N$ 
\begin{equation}
\nu_\mu (k)\, N(p)\, \rightarrow \mu (k^\prime)\, X \ ,
\end{equation}
written in terms of $x=Q^2/(2 p\cdot q)$, $Q^2=-q^2$
$q=k-k^\prime$, $y=p\cdot q/(p\cdot k)$, nucleon mass $M$
is 
\begin{eqnarray}
\label{eqn:dsdxdy} 
\frac{d^2\sigma}{dx\ dy} &&= \frac{2G_F^2 M
E_{\nu}}{\pi(1+Q^2/M_W^2)^2}\nonumber \\
&& \times  \left\{
q(x,Q^2)+(1-y)^2\bar{q}(x,Q^2)\right\} \ .
\end{eqnarray} 
The quantities $q(x,Q^2)$ and $\bar{q}(x,Q^2)$ are the
parton distribution functions (PDFs) describing the quark
and antiquark content of the nucleon.

In Eq. \ref{eqn:dsdxdy}, the high energy behavior of
the cross section can be qualitatively understood.
At low energy, the conventional growth of the cross section
linearly with neutrino energy comes from the fact that at low
$Q^2$, the $W$-boson propagator and the
PDFs are nearly $Q$ independent. Measurements in this linear
regime are made up to energies of approximately 450 GeV \cite{pdg}.

At higher energies, the $Q^2$ dependence becomes important in
two ways. There is a power suppression from the boson propagator,
and there is a logarithmic growth of the PDFs with increasing
$Q^2$. This last feature was
first pointed out in the context of rising neutrino cross sections by
Andreev, Berezinsky and Smirnov in Ref. \cite{abs}.
Qualitatively, the propagator limits the value of $Q^2$ to
approximately
$Q^2\sim M_W^2 \sim 10^4$ GeV$^2$, so for an incident neutrino
energy $E_\nu$, one is probing a range of $x$ values of approximately
\begin{equation}
x\sim \frac{10^4}{(E_\nu/{\rm GeV})}\ .
\end{equation}

At the highest energies discussed here, $E_\nu=10^{12}$ GeV,
this translates to $x\sim 10^{-8}$, well below the measured
region for parton distribution functions at $Q^2\sim M_W^2$ \cite{h1,zeus}.
In the next section, we discuss the parton distribution functions
and their extrapolations to small values of parton $x$. For an extensive and
readable discussion
of deep inelastic scattering and the parton distribution functions,
the reader is referred to Ref. \cite{devcoop}.

\section{PARTON DISTRIBUTION FUNCTIONS AND EXTRAPOLATIONS}

\subsection{DGLAP Evolution}

The cross section for neutrino-nucleon scattering has
been evaluated over the past twenty years with a number
of different input parton distribution functions and
extrapolations to small $x$ values. In the Dokshitzer-Gribov-%
Lipatov-Altarelli-Parisi (DGLAP) formalism \cite{dglap}, valid for large
$Q^2$ and moderate $x$, parton distribution functions of quarks, antiquarks
and gluons are extracted from global analyses and evolved
from a reference $Q_0^2$ to larger values of $Q^2$. 
Typically, modern analyses are performed in the range of 
$10^{-6}-10^{-5}\leq x\leq 1$ with $Q_0^2=1.25-1.69$ GeV$^2$ 
\cite{cteq6,mrst}. 
An alternate approach
of Gluck, Reya and Vogt \cite{grv} is to dynamically generate the
sea quark and gluon distributions from a scale $Q_0^2=0.8$ GeV$^2$
using
DGLAP evolution and requiring consistency with the experimental
data. The GRV PDFs are parametrized between $10^{-9}\leq x\leq 1$.
At $Q\sim M_W$, all of these approaches yied comparable quark
and antiquark PDFs.

Gluon splitting to quark-antiquark pairs,
\begin{equation}
g\rightarrow q\bar{q}
\end{equation}
is responsible for the generation of small $x$ sea distributions,
so by studying $xg(x,Q)$ at small $x$, one gets insight into
the sea distributions as well.
At leading order, it has been shown that for gluon distributions
parametrized at small $x$ by
\begin{equation}
xg(x,Q_0^2)\sim A(Q_0^2)\, x^{-\lambda}\ \quad\quad x \ll 1\ ,
\end{equation}
with moderate values of $\lambda$ ($\lambda\simeq 0.3-0.4$), the gluon
distribution approximately evolves with the same power
law \cite{ekl}
\begin{equation}
xg(x,Q^2)\sim A(Q^2)\, x^{-\lambda}\ .
\end{equation}
Taking this as a guide, one is led to extrapolate sea quark
PDFs below $x_{min}=10^{-6}-10^{-5}$ by matching, e.g.,
\begin{equation}
x\bar{q}(x,Q^2)=\Biggl( \frac{x_{min}}{x}\Biggr)^{\lambda}
x\bar{q}(x_{min},Q^2)
\end{equation}
where $\lambda$ is determined for each flavor from the PDFs at $Q=M_W$.
This is in contrast to the double-logarithmic-approximation
extrapolations \cite{glr} appropriate for
\begin{equation}
xg(x,Q_0^2)\sim {\rm constant}\quad\quad x\ll 1\ 
\end{equation}
which was the standard expectation before HERA results
showed otherwise.

\begin{figure}[htb]
\includegraphics[angle=270,width=15pc]{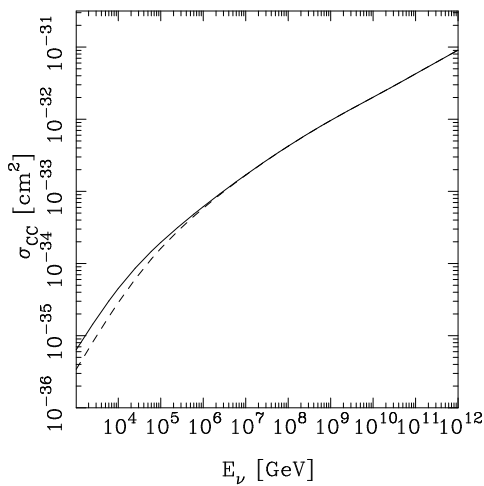}
\caption{The neutrino (solid line) and antineutrino (dashed line) cross section for scattering
with an isoscalar nucleon, computed using the CTEQ6 PDFs [9], extrapolated according to
Eq. (7).}
\label{fig:sig}
\end{figure}

The results of this extrapolation using the NLO CTEQ6 parton
distribution functions \cite{cteq6} 
in the DIS scheme are shown in Fig. \ref{fig:sig}
as a function of incident energy for neutrino
and antineutrino charged current scattering. We use the leading order matrix
element squared to evaluate the cross section. At low
energy, one sees the increase in the cross section proportional
to the neutrino energy, gradually moderated by the $W$-boson
propagator. Above $\sim 10^6$ GeV, the neutrino and antineutrino
cross sections are nearly equal, indicating that the sea quarks
dominate the scattering with neutrinos above that energy.
Other evaluations of the cross sections using a variety of
PDFs, with DGLAP evolution, yield similar results \cite{cs1,cs2,cs3}.

\subsection{BFKL/DGLAP}

\begin{figure}[htb]
\includegraphics[angle=270,width=15pc]{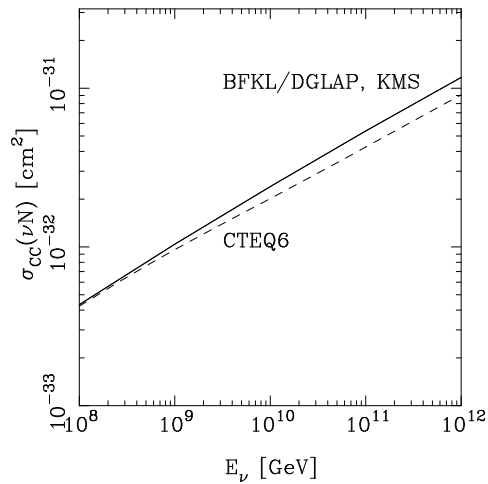}
\caption{The charged current cross section using a power law
extrapolation of the CTEQ6 PDFs and a unified BFKL/DGLAP treatment
of small $x$ by Kwiecinski, Martin and Stasto [18].}
\label{fig:sigcomp}
\end{figure}

At ultrahigh neutrino energies, the values of $x$ probed are
such that $\alpha_s\ln (1/x)$ can be large. The DGLAP formalism
sums powers of $\ln (Q^2/Q_0^2)$ but does not
sum powers of $\ln (1/x)$. Theoretical efforts have been directed
to include summations of $\ln (1/x)$ corrections, the Balitsky,
Fadin, Kuraev and Lipatov (BFKL) formalism \cite{bfkl}. Because of the 
$Q^2$ evolution of the PDFs are so important, only a unified
BFKL/DGLAP approach is useful for UHE neutrinos.
A combined BFKL/DGLAP evaluation of the neutrino-nucleon
cross section has been made
by Kwiecinski, Martin and Stasto (KMS) \cite{kms}.
The approach calls for generalized parton distribution
functions, which, when integrated over parton transverse
momentum, yield the usual PDFs. They take as their canonical
PDFs the GRV parametrization,  and their results for the
charged current cross section are shown by the
dashed line in Fig. \ref{fig:sigcomp}. Also shown in
Fig. \ref{fig:sigcomp} is the CTEQ6
result with a power law
extrapolation for comparison.
The close correspondence of the two curves suggests that
the additional $\ln (1/x)$ corrections are small compared to
the DGLAP evolution. This is consistent with other theoretical
work on unifying the BFKL and DGLAP approaches \cite{altarelli}.

\subsection{Saturation}

The growth of the cross section with energy must eventually saturate
to preserve unitarity \cite{glr,saturation}. In parton model language, saturation comes from
\begin{equation}
gg\rightarrow g\ ,
\end{equation}
gluon recombination. In principle, gluon recombination would come
into the evolution equations via a non-linear term.
A first estimate of the saturation effect is
to consider
\begin{equation}
\frac{\alpha_s}{Q^2} xg(x,Q^2)\sim \pi R^2
\end{equation}
where $\pi R^2$ is the transverse size of the proton disk.
For $Q^2\sim M_W^2$, this lead to an estimate of $x\sim 10^{-17}$
for the relevant scale of $x$ \cite{vogelsang}. 

\begin{figure}[htb]
\includegraphics[angle=270,width=15pc]{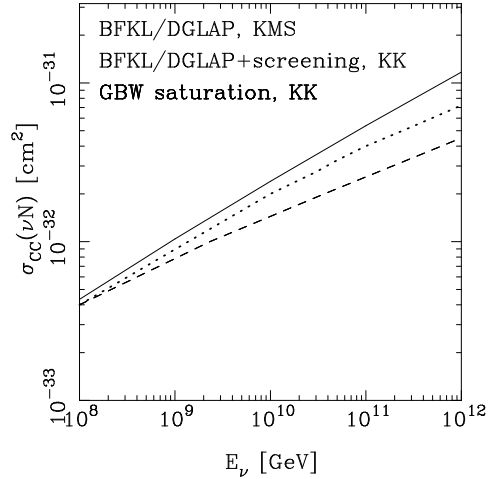}
\caption{Adapted from Ref. [22] of Kutak and Kwiecinski (KK), showing the
KMS result of Ref. [18] for a unified BFKL/DLAP approach (solid),
a unified BFKL/DGLAP approach with screening (dotted) and the evaluation
with a color dipole scattering approach using the Golec-Biernat
and Wusthoff scattering model.}
\label{fig:sigkk}
\end{figure}

Kutak and Kwiecinski (KK) have instead evaluated a unified BFKL/DGLAP
equation with a nonlinear term accounting for gluon recombination.
Their results \cite{kk} together with the Kwiecinski, Martin and
Stasto 
\cite{kms}
curve without recombination are shown in Fig. \ref{fig:sigkk}. Also shown
are Kutak and Kwiecinski's results using  the Golec-Biernat
and Wusthoff (GBW) color dipole model \cite{gbw} as an alternative
to the unified BFKL/DGLAP plus gluon recombination approach.

The cross sections are remarkably consistent at the highest energies.
The lower curve in Fig. \ref{fig:sigkk} is only about a factor of
two lower than the uppermost curve. This is a reasonable estimate of the
range of predictions. Additional work on saturation appears in
Refs. \cite{fiore,machado}, with results  that lie in the same range.
Other authors have suggested that in fact the ultrahigh energy
neutrino cross section is significantly enhanced by QCD effects
\cite{jamal}, 
however, one awaits
a quantitative demonstration of the effect in this context.

\section{NON-PERTURBATIVE AND NON-STANDARD MODEL CONTRIBUTIONS}

\subsection{Non-perturbative electroweak instantons}

Electroweak instanton contributions to the neutrino-nucleon
cross section have been discussed in several references \cite{ew1,ew2}.
The idea is that standard model electroweak instanton processes
can result in neutrino-nucleon conversion to multiparticle final
states. They are exponentially suppressed because it is tunneling
phenomenon with an energy barrier, the sphaleron energy,
equivalent to approximately 8 TeV. Because it is a tunneling
process between two vacua, the transition rate is exponentially
suppressed at low energies, but can become quite large at high
energies.

The actual calculation of the electroweak instanton contribution
to the neutrino-nucleon cross section is subject to large 
uncertainties, both in the energy at which the transition is
unsuppressed, and the normalization of that large cross section.
Two different approaches, one using
a perturbative approach with an instanton background field \cite{ringwald},
and the other using a semi-classical
approach \cite{bezrukov}, give widely different thresholds for the large cross section:
the first at $\sim 30$ TeV neutrino-parton center of mass energy, and the
other at an energy scale roughly 10 times larger. The estimates of the magnitude
of the cross section above the scale at which these interactions
turn on is also widely varying, on the order of three orders of magnitude for the
work of \cite{ew1} and \cite{ew2}.
The ability of neutrino telescopes to constrain instanton induced
events has been explored \cite{ew1,ew2}, with the conclusion that it may be difficult
despite the very large cross sections.

\subsection{Large extra dimensions}

Non-standard model contributions to the neutrino-nucleon
cross section are not well constrained at ultrahigh energies.
Indeed, interactions at the millibarn level may help to explain
the highest energy cosmic ray events by interpreting them
as being produced by neutrinos with anomalously strong interactions.
One model that has an anomalously large neutrino cross section
is the model of $n$ large compact extra dimensions \cite{orig}. Theoretical
analyses have shown that one consequence of large extra dimensions
is the possibility to produce mini-black holes in neutrino-nucleon
interactions \cite{bhprod,bhprod2,bhprod3,bhprod4,domokos}. 
These mini-black holes decay into 
hadrons and leptons \cite{bhdecay},
and so mimic showers produced by cosmic ray interactions in the
atmosphere. 

In these mini-black hole calculations, one takes the neutrino-parton
cross section to be
\begin{equation}
\label{eq:bhsig}
\hat{\sigma}(\nu j\rightarrow BH) = \pi R_s^2\mid_{M_{BH}=\sqrt{\hat{s}}}\theta
(\sqrt{\hat{s}}-M_{BH}^{\rm min})
\end{equation}
where the Schwarzschild radius $R_S$ is given by
\begin{equation}
R_S = \frac{1}{M_D} \Biggl[ \frac{M_{BH}}{M_D}\Biggl( \frac{2^n\pi^{\frac{n-3}{2}}
\Gamma (\frac{3+n}{2})}{2+n} \Biggr)\Biggr]^{\frac{1}{1+n}}\ .
\end{equation}
The geometrical cross section is evaluated for neutrino-parton
center of mass energy $\sqrt{\hat{s}}$ larger than some
minimum black hole mass $M_{BH}^{\rm min}$, which should be larger than the scale of the extra
dimensions $M_D$ so that the semiclassical approach is a reasonable one.
The neutrino-nucleon cross section then involves the integral over parton $x$ 
of the neutrino-parton cross section in Eq. (\ref{eq:bhsig})
multiplied by the parton distribution function.

\begin{figure}[htb]
\includegraphics[angle=270,width=15pc]{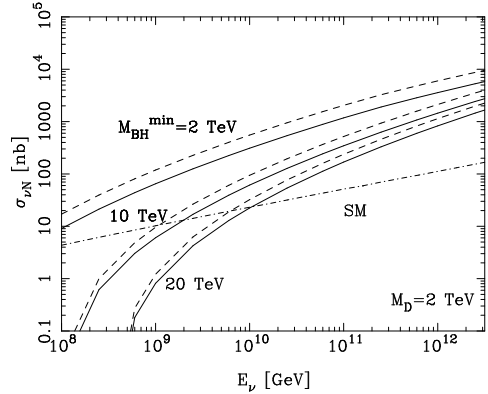}
\caption{The black hole production cross section for neutrino-nucleon scattering
with $n=4$ (solid) and $n=6$ (dashed) extra dimensions for
$M_{BH}^{\rm min}=2,\ 10$ and 20 TeV, given $M_D=2$ TeV. The
standard model cross section is also shown (dot-dashed).}
\label{fig:sigbh}
\end{figure}

An illustration of the enhanced neutrino-nucleon cross section from
mini-black hole production is shown in Fig. \ref{fig:sigbh}. For this figure, the
scale of the extra dimensions is set at $M_D=2$ TeV, and the cross section
is shown for a range of minimum black hole masses, for $n=4$ (solid line) and 6 
(dashed line) extra
dimensions.
The standard model cross section is shown by the dot-dashed line for reference.

There has been an extensive literature on the topic of 
air shower constraints on the parameters $M_D$ and $n$ \cite{bhprod2,bhprod3,bhprod4}.
Ahn, Cavaglia and Olinto \cite{olinto} have recently pointed out, however, that 
by including reasonable estimates of the uncertainties,
the air shower constraints on the mini-black hole parameters
do not exceed the Tevatron Collider limits.

While there are uncertainties in evaluating contributions from
non-standard model physics, ultrahigh energy neutrinos may offer an unparalleled
opportunity to explore particle physics in new energy regimes \cite{review},
whether it is large extra dimensions, supersymmetry or some other extension
of the standard model.

\section{FINAL REMARKS}

Kusenko and Weiler have made the point that the cross section
plays different roles in neutrino air shower events and
upward-going events \cite{kusenko}. As an example, they point to the case
of tau neutrinos. Tau neutrinos can produce horizontal air
showers, but they can also produce upward air showers. The upward air
showers come from the two step process of first $\nu_\tau$ charged
current interaction in the Earth producing a tau, which emerges from
the Earth to decay in the atmosphere. Event rates for
neutrino interactions with air nuclei producing horizontal
air showers increase linearly with the cross section.
On the other hand, the upward air shower rate from tau
decays is suppressed by large cross sections. Indeed, at ultrahigh
energies, the upward air showers will be nearly horizontal, because
neutrino fluxes incident at more than a ``skimming'' angle will
be extinguished due to the short interaction length of the
neutrino \cite{feng}. If ultrahigh energy neutrino fluxes are large
enough, one might get an effective measurement of the
neutrino cross section by comparing these two processes.

Neutrino messengers from astrophysical sources show the promise
of revealing the nature of the sources themselves. Neutrino
detection will give us more information about particle physics
in energy regimes far beyond the reach of terrestrial accelerators,
testing the standard model and extensions to new physics at 
ultrahigh energies.

\end{document}